\documentclass{iau}
\usepackage{graphicx}
\usepackage{amsmath}
\usepackage{comment}
\usepackage{rotating}
\usepackage{color}
\usepackage{amsfonts}

\title[Optimal observables in galaxy surveys] 
{Optimal observables in galaxy surveys}

\author[Julien Carron \& Istv\'an Szapudi]   
{Julien Carron$^1$
 \and Istv\'an Szapudi}

\affiliation{$^1$Institute for Astronomy, University of Hawaii, \\ 2680 Woodlawn Drive,
Honolulu, HI 96822,USA \\ email: {\tt carron@ifa.hawaii.edu} \\[\affilskip]
}

\pubyear{2014}
\volume{306}  
\pagerange{119--126}
\jname{Statistical Challenges in 21st Century Cosmology}
\editors{A.C. Editor, B.D. Editor \& C.E. Editor, eds.}

\newcommand{\beq}{\begin{equation}}
\newcommand{\enq}{\end{equation}}
\newcommand{\veck}{\mathbf k}

\newcommand{\lb}{\left [}
\newcommand{\rb}{\right ]}
\newcommand{\lp}{\left (}
\newcommand{\rp}{\right )}
\newcommand{\av}[1]{\left \langle #1 \right\rangle}

\begin{document}

\maketitle

\begin{abstract}
The sufficient statistics of the one-point probability density function of the dark matter density field is worked out using cosmological perturbation theory and tested to the Millennium simulation density field. The logarithmic transformation is recovered for spectral index close to $-1$ as a special case of the family of power transformations. We then discuss how these transforms should be modified in the case of noisy tracers of the field and focus on the case of Poisson sampling. This gives us optimal local transformations to apply to galaxy survey data prior the extraction of the spectrum in order to capture most efficiently the information encoded in large scale structures.
\end{abstract}

\firstsection 
\section{Motivations}
Among the principal motivations behind this study of the cosmological information within the matter density field are i) the improved quality and size of modern and future cosmological data sets, that should allow understanding of more than the traditional and well-understood yet fairly crude descriptor the power spectrum. Its statistical power is known to be limited beyond the linear regime due to the tri-spectrum including beat-coupling (\cite{RimesHamilton06}), or super-sample covariance (\cite{TakadaHu13}) caused by large scales modes ii) the very specific type of non-Gaussianity induced by gravity, which is characterised by extreme events, that renders mainstream tools designed for mildly non-Gaussian fields, higher order $N$-point functions, inadequate, as showed by \cite{CarronNeyrinck12} iii) the correct generalization of non-linear transformations (such as those of \cite{NeyrinckEtal09} or \cite{SeoEtal11}) to noisy tracers of the fields. This is important as noise modifies the statistical properties of the data and the optimal statistics or transform must take this into account in order to be efficient.
\section{Overview}
\begin{figure}[b]
\begin{center}
\includegraphics[width=0.8\textwidth]{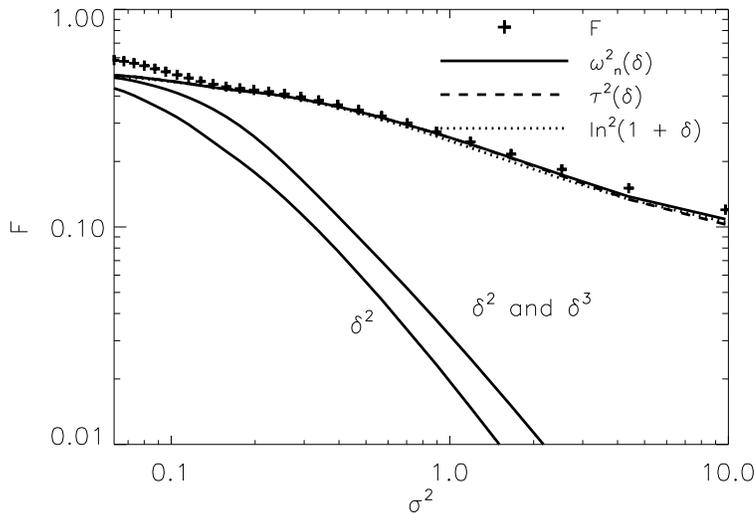} 
 \caption{The information content of various statistics in the Millennium density field simulation for different smoothing scales shown as a function of the variance. The crosses show the total information of the one-point PDF. The moments (lower lines) perform poorly as for lognormal field statistics. The statistics derived in this work (upper lines) are maximally efficient over the full range.}
   \label{figmill}
\end{center}
\end{figure}

We use the fact that for any PDF $p$ and parameter $\alpha$ of interest, the observable defined as $\partial_\alpha \ln p$ always captures the entire Fisher information content $F$ of the PDF. It is therefore a 'sufficient statistics'. Our starting point is the formal Edgeworth series expansion for the logarithm of the PDF (here for one variable),
\beq \label{Edge}
\ln p(\nu) = -\frac{\nu^2}{2} -\frac 12 \ln \sigma^2 + \sum_{n  = 1}^\infty \sigma^n g_{n+2}(\nu), \quad \quad \nu = \frac{\delta}{\sigma}.
\enq
In that equation $g_{k}(\nu)$ is a polynomial of degree $k$ given by combinations of Hermite polynomials and the cumulants. Terms proportional to some power $\nu^k$ of the field enters only with power of $\sigma^{k-2}$ and higher. This simple form of $\ln p$ allows us to see through the structure of the information within the moments of the field : in the expansion of $F$ in powers of $\sigma^2$, it is always possible to capture the $k$ first terms with a polynomial of order $k + 2$. The variance $\delta^2$ captures the leading Gaussian information, a degree three polynomial in the field will capture the next to leading term, etc. This gives us in the next section the Taylor expansion of the sufficient statistic of the PDF. The same structure holds of course for the hierarchy of $N$-point functions associated to the multivariate PDF, giving us the multivariate Taylor expansion of the optimal observables. This is a generalization left for future work.
\newline
\newline
{\underline{\it The sufficient observable of the matter field one-point PDF}}.
\newline
Next we go further with the one-point PDF as a function of scale in details. A useful reference is \cite{CarronSzapudi13b}.  The log-variance $\ln \sigma^2$ can be chosen as the relevant parameter. By reorganizing the series in power of $\delta$ rather than $\nu$, the sufficient statistic has a complicated form involving different functions of $\delta$,
\beq
f_0(\delta) + \sigma^4f_4(\delta) + \sigma^6f_6(\delta) + \cdots,
\enq
Nevertheless, it is found that the leading function $f_0(\delta)$ completely dominates the information. Besides, $f_0(\delta)$ depends only on the leading order cumulants. Writing  $f_0$ as a power series $f_0 = \delta^2 + a_3\delta^3 + \cdots$,  we gather that the coefficients are the leading coefficients of the polynomials $g_k$ in Eq. \eqref{Edge}. These coefficients are explicitly given by
\beq \label{f0}
a_n  =\frac{2}{n!} \sum_{\veck} (-1)^{|\veck| }\lp n -2+ |\veck| \rp! \prod_{i \ge 3}  \frac{S_i^{k_i}}{  \lp i-1 !\rp^{k_i}k_i!},
\enq
where the sum runs over all vectors of positive integers $\veck = (k_3,k_4,\cdots)$ of any dimension such that $\sum_{i} i k_i =  n - 2 $, and where $|\veck |$ stands for $ \sum_i k_i$. Given the well known values of the leading cosmological cumulants calculated by Bernardeau (\cite{Bernardeau94}), we can obtain explicitly the first few coefficients for power-law power spectra $P(k) \propto k^n$. Two simple functions provide an almost perfect match for any value of $n$ of interest. First the square of the power transformation $\omega_n^2$ (Box-Cox transformation)
\beq \label{omegan}
\omega_n(n) = \frac{(1 + \delta)^{(n+1)/3} -1}{(n+1)/3},\quad \tau(n) =  \frac 32 \lp 1 + \delta \rp^{(n+3)/6} \lb (1 + \delta)^{-2/3} -1 \rb.
\enq
and second the squared linear density contrast $\tau^2$ from spherical collapse. Note that in the former case we recover precisely the logarithmic transform  $\ln (1 + \delta)$  of \cite{NeyrinckEtal09} for $n  = -1$.
\newline
\newline
{\underline{\it Test to the Millennium simulation density field}}.
We tested our results and transforms $\omega_n$ and $\tau_n$ using the $\Lambda$CDM, $z = 0$ matter density field from the Millennium simulation by \cite{SpringelEtal05}. We extracted from the $500h^{-1}$ Mpc box  the one-point PDF on scales $i  \times 1.95^{-1}$Mpc, with $i = 1,\cdots,29$, corresponding roughly to $\sigma^2 \sim 10 - 0.1$. Poisson noise is negligible on all these scales. We then obtained straightforwardly the derivatives of the PDF with respect to $\ln \sigma^2$ using finite differences.  The spectral index was estimated according to $n = -3 - \partial_{\ln R} \ln \sigma^2$ at each scale and lies between $-0.8$ and $-1.2$ at all scales. This gives us then both the total information content of the PDF as well as the efficiency of the statistics introduced above. Fig. \ref{figmill} shows $F$ as the crosses, as a function of $\sigma^2$.  The three upper lines almost indistinguishable from $F$ show the efficiency of $\omega_n^2$, $\tau^2$ as well as the logarithmic mapping $\ln^2(1 + \delta)$. They are efficient over the full range. The two lower solid lines show the efficiency of the variance and that of the variance and third moment jointly. They show the very same behavior than in lognormal fields, becoming rather dramatically poor quite quickly. This is not very surprising given the non-analytic form of the three transforms for $n \sim -1$ with Taylor expansions breaking down quickly for moderate values of $\delta$. At this point we can only speculate that the same happens for higher order moments as well, as the finite volume of the simulations did not allow us to obtain the PDF sufficiently accurately for this purpose.
\newline
\newline
{\underline{\it Optimal transforms in the presence of observational noise}}.
We show now how to adapt our statistics in the presence of noise. For galaxies sampling the density field, we can write very generically for the probability of observing numbers $N$ of galaxies in cells (in a one-dimensional notation for clarity)
\beq 
P(N|\alpha) = \int_{-\infty}^\infty\:dA\:p_A(A|\alpha) P(N | A),\quad \textrm{   with } A = \ln (1 + \delta).
\enq
After differentiating under the integral sign and using Bayes theorem the sufficient statistics for $\alpha$ becomes
\beq
\partial_\alpha \ln P(N|\alpha) = \int_{-\infty}^{\infty} dA\:p(A|N)\partial_\alpha \ln p_A(A,\alpha) = \av{\partial_\alpha \ln p_A(A,\alpha)}_{AIN}.
 \enq
Note that the weight function is now the posterior probability for $A$ given the observations $N$. In the case that the data constrains well the signal the sufficient statistics of the observation becomes simply the sufficient statistics of the signal evaluated at its value favored by the data $A^*(N)$. Alternatively one can apply a saddle-point approximation to the above integral effectively treating the posterior as a Gaussian. Due to the above we can safely use a lognormal signal, to which we add Poisson sampling. This gives the following non-linear equation to solve for the saddle point,
\beq
A^*(N)  + \bar N \sigma^2_Ae^{A^*(N)}= \sigma^2_A \lp N -1/2 \rp. 
\enq
Further the mean and variance and $A^*(N)$ capture the entire information of $P(N)$, see \cite{CarronSzapudi14}.
\begin{figure}[h]
\begin{center}
 \includegraphics[width=0.8\textwidth]{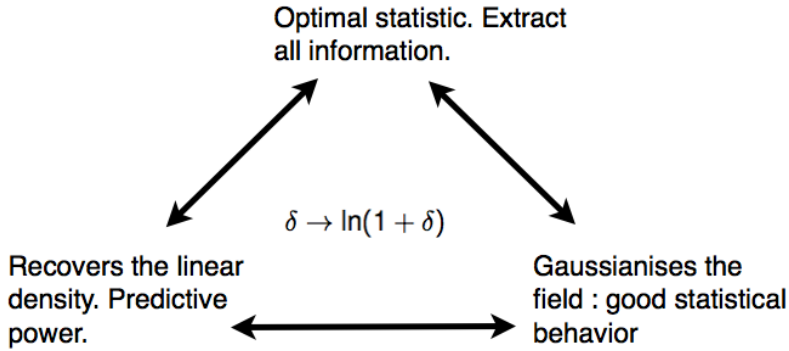} 
 \caption{Three different perspectives on the logarithmic mapping of the matter density field. By way of the arguments of this paper, they are roughly equivalent. This is only because of Gaussian initial conditions linking notably the lower two corners and the fact that $n \sim -1$ on the scales of interest.}
   \label{fig3}
\end{center}
\end{figure}
\section{Discussion}

Our rigorous derivation of the sufficient statistics of the one-point PDF from cosmological perturbation theory points towards
the well-known logarithmic transform as the optimal local transformation. In fact, this work unifies three different facets of that particular transform, illustrated in Fig. \ref{fig3} : the capture of the entire information (because $n \sim -1$), the undoing of the non-linear dynamics and the Gaussianization of the PDF (because of the Gaussian initial conditions). One key aspect of the methods introduced is the systematic way the sufficient statistics and transforms are adapted to the presence of noise. The use of non-linear transforms was until now mostly useful in cosmology in simulations of noise-free fields. This opens the door to the analysis of actual data with efficient non-linear transformations. The analysis of the projected Canada-France-Hawaii Telescope Large Survey (CFHTLS\footnote{\texttt{http://www.cfht.hawaii.edu/Science/CFHLS/}}) data using the power spectrum of the $A^*$ non-linear transform will be exposed by \cite{WolkEtal14}.

\end{document}